\documentclass[11pt]{article}
\usepackage[dvipdfmx]{graphicx}
\usepackage{url}
\usepackage{typearea}
\typearea{20}

\providecommand{\keywords}[1]
{
  \small	
  \textbf{\textit{Keywords---}} #1
}

\begin{document}

\title{Proof of Team Sprint: A Collaborative Consensus Algorithm for Reducing Energy Consumption in Blockchain Systems}
\author{Naoki Yonezawa \thanks{\texttt{n.yonezawa@thu.ac.jp}. Faculty of Humanities and Social Sciences, Teikyo Heisei University.}}
\date{}
\maketitle

\begin{abstract}
This paper introduces Proof of Team Sprint (PoTS), a novel consensus algorithm designed to address the significant energy inefficiencies inherent in traditional Proof of Work (PoW) systems. PoTS shifts the consensus mechanism from an individual competition model to a collaborative team-based approach. Participants are organized into groups, with each group collaboratively working to solve cryptographic puzzles required to validate transactions and add new blocks to the blockchain. This collaborative approach significantly reduces the overall energy consumption of the network while maintaining high levels of security and decentralization. Our analysis shows that PoTS can reduce energy consumption by a factor of $1/N$, where $N$ is the number of participants in each group, compared to PoW. Furthermore, PoTS maintains a fair and equitable reward distribution among participants, ensuring continued engagement and network integrity. The paper also discusses the scalability, security implications, and potential challenges of adopting PoTS, positioning it as a promising alternative for sustainable blockchain technology.
\end{abstract}


\section{Introduction}

\subsection{Background}
\subsubsection{Overview of Blockchain Technology}
Blockchain technology, since its inception with Bitcoin~\cite{nakamoto2008peer} in 2008, has emerged as a groundbreaking innovation that fundamentally transforms the way digital transactions are conducted. At its core, a blockchain is a distributed ledger that records transactions across a network of computers in a secure, transparent, and immutable manner. Each block in the chain contains a list of transactions, and once a block is validated by the network, it is added to the existing chain, making it virtually impossible to alter any information without altering all subsequent blocks---a feature that ensures the integrity and security of the data.

The decentralized nature of blockchain eliminates the need for intermediaries, such as banks or financial institutions, by enabling peer-to-peer transactions directly between parties. This not only reduces transaction costs but also increases the speed and efficiency of operations across various industries, including finance, supply chain management, healthcare, and beyond.

Blockchain's potential extends far beyond cryptocurrencies. It has enabled the development of decentralized applications (dApps)~\cite{cai2018decentralized}, smart contracts~\cite{wang2018overview,mohanta2018overview}, and tokenization of assets~\cite{roth2021tokenization}, contributing to the growth of decentralized finance (DeFi)~\cite{schar2021decentralized} and non-fungible tokens (NFTs)~\cite{wang2021non}. These applications demonstrate blockchain's versatility and its ability to disrupt traditional business models by providing new ways of conducting transactions and managing digital assets.

\subsubsection{The Role of Consensus Algorithms in Blockchain}
At the heart of blockchain technology lies the consensus algorithm---a critical component that ensures all participants in the network agree on the validity of transactions and the state of the blockchain. Consensus algorithms are responsible for maintaining the security, decentralization, and integrity of the blockchain, even in the presence of malicious actors.

The most widely known and implemented consensus algorithm is Proof of Work (PoW)\cite{back2002hashcash}, which was introduced by Bitcoin and has been foundational to the security model of many blockchains. In PoW, network participants, known as miners, compete to solve complex cryptographic puzzles. The first miner to solve the puzzle gets the right to add a new block to the blockchain and is rewarded with newly minted coins and transaction fees. This process, while highly secure, is also computationally intensive and energy-consuming, leading to significant environmental concerns and prompting the search for more sustainable alternatives.

Over time, various alternative consensus algorithms have been developed to address the limitations of PoW. Proof of Stake (PoS)~\cite{king2012ppcoin}, for example, replaces the energy-intensive puzzle-solving process with a system where validators are chosen to create new blocks based on the number of coins they hold and are willing to ``stake'' as collateral. Other algorithms, such as Delegated Proof of Stake (DPoS)~\cite{9617549} and Byzantine Fault Tolerance (BFT)~\cite{lamport2019byzantine}, have also emerged, each with unique approaches to achieving consensus in a decentralized network.

Despite the evolution of these algorithms, the challenge of balancing security, scalability, and sustainability remains. This ongoing quest has led to the exploration of collaborative consensus mechanisms, where the focus shifts from individual competition to teamwork and shared incentives. It is within this context that the concept of Proof of Team Sprint (PoTS) is introduced, aiming to reduce energy consumption while maintaining the security and integrity of the blockchain.

\subsection{Motivation}
\subsubsection{Energy Consumption in Proof of Work (PoW)}
PoW, while foundational to the security and decentralization of many blockchain systems, has become increasingly scrutinized due to its immense energy consumption. The PoW mechanism requires miners to solve complex cryptographic puzzles to validate transactions and add new blocks to the blockchain. This process, often described as a computational race, involves a vast amount of trial-and-error calculations, with only the first miner to find a valid solution being rewarded.

The competitive nature of PoW means that miners are incentivized to continually invest in more powerful hardware and consume more energy to increase their chances of winning the race. As a result, PoW-based blockchains, particularly Bitcoin, have been criticized for their environmental impact. 
Recent estimates as of September 2024 suggest that the energy consumption of the Bitcoin network alone rivals that of entire countries, such as Poland or Malaysia.
This level of energy usage not only raises concerns about the sustainability of PoW-based systems but also highlights the broader environmental costs associated with maintaining these decentralized networks.

The carbon footprint of PoW is further exacerbated by the geographical concentration of mining operations in regions where electricity is cheap but often generated from non-renewable sources. This has led to growing pressure from environmental groups, policymakers, and the public to seek more sustainable alternatives that can reduce the energy burden of blockchain technologies without compromising their security.

\subsubsection{The Need for Energy-Efficient Alternatives}
The substantial energy consumption associated with PoW has catalyzed the search for alternative consensus mechanisms that can maintain the security and decentralization of blockchain networks while drastically reducing their environmental impact. As the demand for blockchain technology continues to grow across various industries, from finance to supply chain management, the importance of developing energy-efficient solutions becomes even more critical.

Energy-efficient alternatives to PoW, such as PoS, have emerged as promising solutions. PoS reduces the need for energy-intensive computations by selecting validators based on the number of coins they hold and are willing to ``stake'' as collateral. This approach significantly lowers the computational requirements, leading to a reduction in energy consumption. However, while PoS and its variants address some of the energy concerns, they also introduce new challenges, such as potential centralization risks and the ``nothing at stake'' problem.

In response to these challenges, there is a growing interest in exploring consensus mechanisms that not only reduce energy consumption but also promote collaboration and fairness within the network. PoTS is proposed as a novel approach that leverages collaborative efforts among participants to achieve consensus. By shifting the focus from individual competition to team-based cooperation, PoTS aims to distribute the computational load more efficiently and minimize redundant calculations, thereby significantly reducing the overall energy consumption of the network.

The development of PoTS is motivated by the need to create a more sustainable and scalable blockchain ecosystem. As the world increasingly prioritizes environmental responsibility, the adoption of energy-efficient consensus mechanisms like PoTS could play a crucial role in the future of blockchain technology, enabling its continued growth and integration into a wide range of applications without exacerbating the global energy crisis.

\subsection{Contribution of This Paper}
\subsubsection{Introduction of Proof of Team Sprint (PoTS)}
This paper introduces Proof of Team Sprint (PoTS), a novel consensus algorithm designed to address the energy inefficiencies associated with traditional PoW systems. PoTS leverages the collaborative efforts of multiple participants within a blockchain network to achieve consensus through a team-based approach. Unlike PoW, where individual miners compete against each other, PoTS organizes participants into groups that work together to solve cryptographic puzzles. The algorithm distributes computational tasks across team members, allowing for more efficient processing while maintaining the security and decentralization that are critical to blockchain technology.

The introduction of PoTS represents a significant departure from the competition-driven model of PoW, emphasizing cooperation and shared incentives instead. By fostering collaboration, PoTS not only reduces the computational redundancy inherent in PoW but also creates an environment where the energy required to maintain the network is significantly lowered. This paper outlines the fundamental principles of PoTS, detailing its design, operation, and potential advantages over existing consensus mechanisms.

\subsubsection{Comparative Analysis with PoW}
A key contribution of this paper is the comparative analysis between PoTS and PoW. While PoW has been instrumental in establishing the security and trustworthiness of blockchain systems, its reliance on massive computational power has raised concerns about its scalability and environmental sustainability. In contrast, PoTS offers a more energy-efficient alternative by restructuring how consensus is achieved.

This paper contrasts PoTS with PoW on several fronts, including computational efficiency, energy consumption, security, and scalability. By examining these aspects, we demonstrate how PoTS can achieve similar or superior levels of security while drastically reducing the energy requirements. The analysis also highlights how PoTS mitigates some of the key drawbacks of PoW, such as the centralization of mining power and the environmental impact, without sacrificing the decentralized nature of the blockchain.

\subsubsection{Evaluation from the Perspective of Energy Consumption}
The energy consumption associated with blockchain networks is a critical issue that has gained increasing attention from both the academic community and the general public. As blockchain technology continues to expand its influence across various industries, the need for sustainable and energy-efficient solutions becomes paramount. This paper evaluates PoTS from the perspective of energy consumption, presenting it as a viable alternative to energy-intensive PoW-based systems.

The evaluation is based on a detailed analysis of the energy dynamics within PoTS, including the reduction in redundant computations and the efficient distribution of computational tasks across team members. We provide quantitative estimates of the energy savings that PoTS can achieve compared to PoW, supported by theoretical models.
This assessment underscores the potential of PoTS to significantly lower the environmental footprint of blockchain networks, making it a promising candidate for future blockchain implementations that prioritize sustainability.

Through this evaluation, we aim to establish PoTS as a consensus mechanism that not only meets the security and decentralization standards expected of modern blockchains but also aligns with global efforts to reduce energy consumption and mitigate climate change.

This paper is organized as follows: Section 2 provides a comprehensive overview of the background and related work, focusing on existing consensus mechanisms and their limitations. Section 3 introduces the PoTS algorithm, detailing its design principles and operational mechanics. In Section 4, we conduct an in-depth energy consumption analysis, comparing PoTS with traditional consensus methods. Section 5 discusses the implications of PoTS for blockchain technology, addressing potential challenges. Finally, Section 6 concludes the paper by summarizing the key findings, outlining the potential impact of PoTS on the evolution of decentralized systems, and suggesting directions for future research.

\section{Background and Related Work}
\subsection{PoW and Its Limitations}

\subsubsection{PoW Algorithm Overview}
The PoW algorithm is fundamental to the functioning of many blockchain systems, most notably Bitcoin. An overview of the basic principles underlying PoW explains how it achieves consensus across a decentralized network. PoW requires participants, known as miners, to solve complex cryptographic puzzles in order to validate transactions and add them to the blockchain. The process is designed to be computationally intensive, ensuring that the work required to validate a block is significant enough to deter fraudulent activity. The successful miner, who is the first to solve the puzzle, is rewarded with cryptocurrency, incentivizing participation and maintaining network security. This combination of incentives and security has made PoW instrumental in the widespread adoption of blockchain technology by ensuring the integrity and reliability of decentralized systems. As a result, PoW has become the standard against which other consensus mechanisms are measured, despite its challenges.

\subsubsection{Energy Consumption in PoW}
One of the primary criticisms of the PoW algorithm is its substantial energy consumption. The computational power required to solve cryptographic puzzles leads to high electricity usage. The massive scale of mining operations has resulted in significant environmental concerns, as the energy consumed by the global network of miners continues to grow. This energy consumption is not only costly but also raises questions about the long-term sustainability of PoW as a consensus mechanism. The environmental impact, particularly in terms of carbon emissions, has sparked debates on the viability of PoW in a world increasingly focused on reducing energy use and mitigating climate change. Consequently, the high energy demands of PoW have prompted the search for alternative consensus mechanisms that can achieve similar levels of security and decentralization without the associated environmental costs.

\subsubsection{Scalability and Environmental Impact}
The scalability of the PoW algorithm presents another significant challenge that impacts its environmental footprint. As blockchain networks expand, the demand for faster transaction processing increases, yet PoW struggles to scale efficiently. The inherent limitations of PoW, such as its reliance on intensive computation, restrict its ability to process transactions quickly as network size grows. The need for more computational power not only exacerbates energy consumption but also contributes to slower transaction times, making PoW less effective for large-scale applications. Moreover, as the network expands and more energy is consumed, the environmental impact becomes more pronounced, raising concerns about the ecological sustainability of PoW. These challenges underscore the need for alternative solutions that can provide the security and decentralization benefits of PoW while being more scalable and environmentally sustainable.

\subsection{Related Work}
In response to the significant energy consumption and scalability issues associated with PoW, several alternative consensus algorithms have been proposed. Among these, PoS~\cite{king2012ppcoin} has emerged as a leading contender due to its energy-efficient design. Unlike PoW, which requires miners to perform computationally intensive tasks, PoS selects validators based on the number of coins they hold and are willing to ``stake'' as collateral. This approach drastically reduces the energy required to secure the network while still maintaining a high level of security. Validators in PoS are incentivized to act honestly, as they stand to lose their staked assets if they attempt to validate fraudulent transactions. PoS has been adopted by several blockchain platforms, including Ethereum 2.0, highlighting its growing acceptance. However, PoS is not without its limitations, such as the potential for centralization if a small number of participants control a large portion of the stake.

Building on the PoS model, DPoS~\cite{9617549} introduces a more efficient consensus mechanism by allowing stakeholders to vote for a small number of delegates who are responsible for validating transactions and maintaining the blockchain. This system significantly improves transaction throughput and reduces latency, making it more suitable for high-frequency applications. DPoS also addresses some of the centralization concerns of PoS by distributing power among a group of elected delegates rather than concentrating it in the hands of a few large stakeholders. However, DPoS introduces its own challenges, such as the risk of delegate collusion and the potential for reduced decentralization if voter participation is low. Despite these challenges, DPoS has been successfully implemented in platforms like EOS and BitShares, where its efficiency and scalability are key advantages.

Another important class of consensus algorithms is BFT~\cite{lamport2019byzantine} and its various adaptations. BFT is designed to ensure that a distributed system can reach consensus even in the presence of malicious actors or nodes that fail to behave correctly. By requiring agreement among a majority of nodes, BFT algorithms can maintain the integrity of the system even when some participants attempt to disrupt the process. Variants of BFT, such as Practical Byzantine Fault Tolerance (PBFT)~\cite{castro1999practical}, have been developed to improve scalability and performance, making BFT a viable option for permissioned blockchains and other environments where high security is critical. However, BFT and its variants often face scalability issues in large, decentralized networks due to the high communication overhead required for consensus. Despite these challenges, BFT remains a foundational element in the design of secure and reliable distributed systems.

FairCoin~\cite{konig2018proof} employs another cooperative approach known as Proof of Cooperation (PoC). In PoC, blocks are created not through competition, but by collaboration among Cooperatively Validated Nodes (CVNs). These nodes, verified through a peer consensus process, take turns in creating blocks, ensuring energy-efficient and fair block generation. Unlike PoTS, which emphasizes teamwork in solving cryptographic puzzles, PoC operates through a more structured round-robin method where cooperation is predefined and trusted participants take on specific roles. While both PoTS and PoC aim to reduce energy consumption by promoting collaboration, PoC focuses more on a stable, cooperative network of trusted nodes rather than the dynamic teamwork seen in PoTS.

In addition to these mechanisms, there are over 100 different consensus algorithms in existence~\cite{9376868}, each offering unique solutions to the trade-offs between security, scalability, and energy efficiency. These include algorithms like Proof of Authority, Federated Byzantine Agreement, and Tendermint, demonstrating the diversity and ongoing innovation in consensus mechanisms.

\subsection{The Need for Collaborative Consensus Mechanisms}
As blockchain technology evolves, the limitations of existing consensus mechanisms like PoW and PoS have become increasingly apparent. One promising solution to these challenges is the adoption of collaborative consensus mechanisms, which emphasize cooperation among network participants rather than competition. By enabling participants to work together, these mechanisms can more effectively distribute computational load, leading to significant improvements in energy efficiency. Collaborative mechanisms also have the potential to enhance security by making it more difficult for malicious actors to gain control over the network. Instead of relying on a single participant's computational power or stake, collaborative consensus requires the collective effort of a group, thereby reducing the risk of centralized attacks. These advantages suggest that collaborative consensus mechanisms could address some of the most pressing issues facing current blockchain technologies, making them an essential area of exploration for the future of decentralized systems.

Over the years, various collaborative consensus mechanisms have been proposed and implemented, each with its own set of strengths and challenges. For instance, federated consensus models, like FedChain~\cite{nguyen2023fedchain}, involve selected groups of nodes working together to validate transactions, which improves scalability and reduces energy consumption. Sharding and hybrid consensus models, as seen in Ethereum 2.0, combine competitive and collaborative elements to enhance efficiency and security. Despite significant progress, these models still face challenges such as coordination overhead and maintaining decentralization, underscoring the need for ongoing research and refinement.

The development of PoTS represents a new step in this ongoing evolution. By learning from the successes and shortcomings of previous approaches, PoTS aims to leverage the strengths of collaboration while addressing the challenges that have hindered other models. Through its innovative design, PoTS seeks to overcome the coordination and scalability issues that have plagued earlier collaborative mechanisms, offering a more robust and efficient solution for the future of blockchain technology. As such, the exploration of collaborative consensus mechanisms remains a critical area of research, with PoTS poised to contribute meaningfully to this field.

\section{Proof of Team Sprint (PoTS) Algorithm}
\subsection{Concept and Design Goals}
\subsubsection{Overview of PoTS}
Proof of Team Sprint (PoTS) is a novel consensus algorithm designed to address the significant energy inefficiencies inherent in traditional PoW systems. At its core, PoTS shifts the consensus mechanism from an individual competition model to a collaborative team-based approach. In PoTS, participants in the blockchain network are grouped into teams, where each team works collectively to solve cryptographic puzzles that validate transactions and add new blocks to the blockchain.

The key innovation of PoTS lies in its ability to distribute the computational workload across multiple participants within a team. By organizing participants into teams, PoTS reduces the redundancy of calculations that characterizes PoW, where each miner independently attempts to solve the same problem. Instead, in PoTS, the computational tasks are divided among team members, each contributing to the overall solution in a sequential manner. The first team to successfully complete the task is rewarded, and their block is added to the blockchain.

This collaborative approach not only decreases the total energy consumption required to secure the network but also fosters a more inclusive and decentralized environment. PoTS is particularly well-suited for blockchain systems that prioritize sustainability and efficiency without compromising on security. The algorithm is designed to be adaptable, allowing for flexibility in team formation and task distribution, which can be tailored to the specific needs and scale of the blockchain network.

\subsubsection{Design Principles and Objectives}
The design of PoTS is guided by several key principles and objectives, which are aimed at creating a consensus mechanism that is both energy-efficient and secure while maintaining the decentralized ethos of blockchain technology.

\begin{enumerate}
\item \textbf{Energy Efficiency:}

The primary objective of PoTS is to drastically reduce the energy consumption associated with consensus mechanisms. By leveraging team-based collaboration, PoTS minimizes redundant computations, which are a significant source of energy waste in PoW. The algorithm distributes the computational load among team members, ensuring that energy is utilized more effectively and that overall consumption is reduced.

\item \textbf{Security and Integrity:}

PoTS is designed to maintain the high security standards expected of blockchain networks. The collaborative nature of the algorithm introduces multiple layers of validation, as each team member's contribution is dependent on the previous member's output. This sequential dependency helps to secure the network against common attacks, such as double-spending and collusion. Additionally, the randomness in team formation reduces the risk of centralization and ensures that no single entity can easily dominate the network.

\item \textbf{Decentralization and Fairness:}

Decentralization remains a cornerstone of PoTS. The algorithm is designed to prevent the concentration of power by ensuring that all participants, regardless of their computational resources, have an equal opportunity to contribute and earn rewards. The random selection of team members and the equitable distribution of tasks promote fairness and inclusivity within the network, making it accessible to a wider range of participants.

\item \textbf{Scalability and Adaptability:}

PoTS is built to be scalable, capable of supporting a growing number of participants and transactions without compromising performance. The team-based structure allows for flexible adjustment of team sizes and task complexity, making it adaptable to various network conditions and requirements. This scalability ensures that PoTS can be implemented in both small and large blockchain networks, maintaining efficiency and security at any scale.

\item \textbf{Incentive Alignment:}

The incentive structure in PoTS is carefully designed to encourage cooperation rather than competition. By rewarding entire teams for successful block validation, PoTS aligns the interests of individual participants with the health and security of the network as a whole. This cooperative incentive model reduces the likelihood of selfish behavior and promotes a more harmonious and sustainable network environment.

\end{enumerate}

In summary, PoTS represents a significant evolution in consensus algorithms, offering a sustainable, secure, and decentralized alternative to traditional models. Its design principles and objectives are centered on reducing energy consumption, enhancing security, and promoting fairness, making it a compelling choice for the next generation of blockchain systems.

\subsection{Detailed Description of the PoTS Algorithm}
\subsubsection{Group Formation and Random Selection of Participants}
The foundation of the PoTS algorithm lies in its unique approach to group formation and the random selection of participants. This process is critical to ensuring that the network remains decentralized, secure, and efficient. In PoTS, participants are organized into teams, each consisting of a predetermined number of nodes ($N$). These teams work together to solve cryptographic puzzles and validate blocks, with the first team to successfully complete the task being rewarded. The formation of these teams and the random selection of participants are essential to maintaining the integrity and fairness of the network.

\paragraph{Selection Process}
The selection process in PoTS is carefully designed to be both fair and unpredictable, ensuring that no single participant or group of participants can dominate the network. The process begins with the identification of eligible nodes within the network. To be eligible, nodes must meet certain criteria, such as maintaining a minimum level of participation, computational power, and network connectivity. This ensures that only active and capable nodes are selected for participation in the consensus process.

In PoTS, the selection of participants is achieved using a novel approach that leverages the blockchain infrastructure itself. A large random number is first generated and recorded on the blockchain. This random number serves as the basis for selecting a smaller set of nodes from the network. Each node applies the same deterministic algorithm (such as taking the remainder when divided by a specific integer) to the random number, which results in the selection of a subset of nodes.

The selected nodes then generate additional random numbers and broadcast them to the network. These random numbers are then combined using a bitwise XOR operation, resulting in a single large random number. This final random number is used as a seed for the random selection algorithm, for instance, by using Python's \texttt{random.seed()} function. With this seed, the network can use \texttt{random.choices()} to form groups of $N$ nodes without requiring further communication.

To ensure the continuity and integrity of the selection process, the newly generated random number is also recorded on the blockchain. This recorded number serves as a permanent record and is used as the basis for the next round of random number generation. By chaining the random number generation in this manner, PoTS ensures that the selection process remains transparent, verifiable, and resistant to manipulation over time.

The randomness of this selection process is crucial for preventing any predictable patterns or biases that could be exploited by malicious actors. By using a combination of blockchain-stored random numbers and decentralized random number generation, PoTS ensures that teams are formed in a manner that is both unpredictable and evenly distributed across the network. This approach is resistant to manipulation, as the random number generation is transparent and verifiable by all participants.

The size of each team ($N$) is a configurable parameter that can be adjusted based on the network's size and specific requirements. A larger team size can provide increased security and computational power, while a smaller team size may improve efficiency and speed. The flexibility of the selection process allows the network to adapt to changing conditions while maintaining its overall integrity, ensuring that the consensus mechanism remains robust and resilient.

\paragraph{Ensuring Fairness and Security}
Ensuring fairness and security during the group formation and selection process is paramount to the success of the PoTS algorithm. Several mechanisms are incorporated into the selection process to achieve these goals.
\begin{enumerate}
\item \textbf{Randomness and Unpredictability:}

The use of cryptographic randomness ensures that the selection of participants is truly unpredictable. This randomness is critical for preventing any entity from gaining undue influence over the group formation process. By making the selection process opaque and resistant to prediction, PoTS minimizes the risk of collusion or manipulation by powerful actors.
\item \textbf{Equal Opportunity:}

To maintain fairness, PoTS ensures that all eligible nodes have an equal opportunity to be selected for participation in the consensus process. This is achieved by using a random selection algorithm that does not favor any particular node based on its past performance or computational power. Every eligible node, regardless of its history or capabilities, has an equal chance of being selected and contributing to the network's security and operation.
\item \textbf{Resistance to Sybil Attacks:}

One of the primary security concerns in decentralized networks is the risk of Sybil attacks, where an attacker attempts to overwhelm the network by creating multiple fake identities (Sybil nodes). PoTS mitigates this risk by implementing strict eligibility criteria for nodes. These criteria, such as proof of resource ownership, participation history, and reputation scores, make it difficult for an attacker to introduce a large number of Sybil nodes into the network. Additionally, the random selection process ensures that even if Sybil nodes are introduced, they are unlikely to be grouped together in a way that could compromise the network.
\item \textbf{Dynamic Team Reformation:}

To further enhance security, PoTS incorporates a mechanism for dynamic team reformation. After each consensus round, the teams are dissolved and new teams are formed using the random selection process. This continuous reshuffling of participants ensures that no team remains static, reducing the risk of long-term collusion among participants. This dynamic reformation also helps to distribute the computational load more evenly across the network, preventing any single node or group of nodes from becoming a bottleneck.

\item \textbf{Transparency and Auditability:}

While the selection process is random, it is also designed to be transparent and auditable. The randomness is verifiable by all participants, allowing the network to confirm that the selection process was conducted fairly. This transparency builds trust among participants and enhances the overall security of the network.
\end{enumerate}

In summary, the group formation and random selection of participants in PoTS are carefully designed to balance fairness and security. By ensuring that teams are formed randomly and unpredictably, and by incorporating mechanisms to resist manipulation and attacks, PoTS creates a robust and secure foundation for consensus in a decentralized environment.

\subsubsection{Sequential Hash Calculation Mechanism}
The Sequential Hash Calculation Mechanism is a core component of the PoTS algorithm. This mechanism enables teams of participants to collaboratively solve cryptographic puzzles in a sequential manner, ensuring that each participant's contribution is integral to the overall solution. By structuring the hash calculation process in a sequential and interdependent fashion, PoTS achieves a balance between security, efficiency, and collaboration.

\paragraph{Hashing Process for Each Participant}
In the PoTS algorithm, the hashing process is divided among multiple participants within a team, with each participant responsible for a specific stage of the computation. The process begins with the first participant ($\mathrm{P}_1$) in the team, who receives the initial input data, which typically includes the transaction data to be validated and any relevant metadata, such as the previous block's hash and a team-specific identifier.

\begin{enumerate}
\item \textbf{Initial Hash Calculation by $\mathrm{P}_1$:}

The first participant, $\mathrm{P}_1$, uses the input data to compute the first hash, denoted as $\mathrm{H}_1$. This hash must meet a predefined difficulty level, typically involving a certain number of leading zeros. $\mathrm{P}_1$ iteratively adjusts a nonce value until a valid $\mathrm{H}_1$ is found that satisfies the difficulty requirement.

\item \textbf{Passing $\mathrm{H}_1$ to the Next Participant ($\mathrm{P}_2$):}

Once $\mathrm{P}_1$ successfully computes $\mathrm{H}_1$, this hash is passed to the next participant, $\mathrm{P}_2$. $\mathrm{P}_2$ then takes $\mathrm{H}_1$ as input, along with additional data (e.g., a new nonce or other relevant information), to calculate the next hash, $\mathrm{H}_2$.

\item \textbf{Sequential Hash Calculations by Subsequent Participants:}

This process continues sequentially, with each participant ($\mathrm{P}_3, \mathrm{P}_4, \dots, \mathrm{P}_N$) receiving the hash from the previous participant and using it as the input to compute the next hash. Each hash calculation must independently satisfy the required difficulty level, ensuring that each participant contributes meaningfully to the overall computation.

\item \textbf{Final Hash and Block Validation:}

The final participant in the sequence, $\mathrm{P}_N$, computes the last hash, $\mathrm{H}_N$, which is then used as the basis for validating the block. If $\mathrm{H}_N$ meets the network's criteria for a valid block, the entire team is considered to have successfully completed the puzzle, and the block is added to the blockchain.

\end{enumerate}

This sequential approach ensures that each participant's work is dependent on the previous participant's output, creating a chain of interdependent calculations that must be completed in the correct order for the team to succeed.

\paragraph{Dependencies Between Hash Calculations}
The interdependencies between hash calculations in PoTS are crucial for maintaining the security and integrity of the consensus process. These dependencies ensure that the work performed by each participant is both necessary and verifiable by the rest of the team, preventing any single participant from dominating the process or bypassing the required computations.

\begin{enumerate}
\item \textbf{Sequential Dependency:}

Each hash calculation in the sequence is directly dependent on the previous hash. This means that the output of $\mathrm{P}_1 (\mathrm{H}_1)$ is the input for $\mathrm{P}_2$, and so on. Without a valid $\mathrm{H}_1$, $\mathrm{P}_2$ cannot proceed to compute $\mathrm{H}_2$, and this dependency continues throughout the sequence. This structure ensures that the team must work together in a coordinated manner, with each participant's contribution being essential for the completion of the task.

\item \textbf{Verification of Intermediate Results:}

The sequential nature of the process allows for continuous verification of each participant's work. As each hash is calculated, it can be immediately verified by the other team members and, potentially, by external observers. This real-time verification adds an additional layer of security, as any errors or malicious attempts to alter the computation can be detected and addressed promptly.

\item \textbf{Difficulty Adjustment and Security:}

The difficulty level for each hash calculation can be dynamically adjusted based on the network's overall security requirements. This flexibility allows PoTS to maintain a high level of security by ensuring that each stage of the sequential process is sufficiently challenging. The interdependency also makes it difficult for an attacker to manipulate the process, as they would need to control multiple participants in the sequence and successfully compute all intermediate hashes without detection.

\item \textbf{Fault Tolerance and Redundancy:}

In the event that a participant fails to produce a valid hash (due to computational errors, network issues, or other disruptions), the team can incorporate fault tolerance mechanisms. For example, a backup participant or a re-run of the computation by the remaining team members can be triggered to ensure that the sequence is completed. This redundancy helps to maintain the robustness and reliability of the consensus process.

\item \textbf{Incentive Alignment through Interdependency:}

The interdependent nature of the hash calculations aligns the incentives of all team members. Since the success of the entire team depends on each participant's contribution, there is a strong motivation for all members to perform their calculations accurately and efficiently. This reduces the likelihood of selfish behavior, such as one participant attempting to undermine the process for personal gain.

\end{enumerate}
In summary, the Sequential Hash Calculation Mechanism in PoTS is designed to foster collaboration among participants while maintaining a high level of security. The dependencies between hash calculations ensure that each participant plays a vital role in the consensus process, creating a robust and efficient system for validating transactions and securing the blockchain.

\subsubsection{Group Competition and Reward Distribution}
\paragraph{Mechanism for Identifying the Winning Group}
In PoTS, identifying the winning group is a critical process that ensures the fairness and integrity of the consensus mechanism. This section explains how PoTS evaluates the progress of each group in real-time and determines which group completes the cryptographic computation first. The mechanism involves continuous monitoring of the groups' computational efforts, with a particular focus on the efficiency and accuracy of their calculations. Once a group successfully completes the required computation, it is promptly recognized as the winner. This identification process is designed to be transparent, ensuring that all participants can verify the legitimacy of the results. Moreover, the mechanism is robust against manipulation, preventing any group from unfairly influencing the outcome. The fairness and transparency of this process are essential to maintaining the trust and reliability of the PoTS system.

\paragraph{Reward Distribution and Incentive Structure}
The distribution of rewards and the design of the incentive structure in PoTS play a pivotal role in promoting cooperation among participants. After a group is identified as the winner, the rewards are distributed equally among all members of that group, reflecting their collective effort. This equitable distribution ensures that each participant receives a fair share of the reward, reinforcing the collaborative nature of PoTS. Additionally, the incentive structure is carefully designed to motivate participants to contribute actively to the group's success. By ensuring that all members benefit from the group's achievements, PoTS encourages sustained engagement and discourages behaviors such as freeloading or minimal participation. This incentive system is crucial for maintaining the overall health and sustainability of the network, as it fosters a cooperative environment where participants are incentivized to work together toward common goals.

\subsubsection{Security Considerations in PoTS}
\paragraph{Defense Against Common Attacks}
Security is a critical aspect of any consensus mechanism, and PoTS is designed with robust defenses against common attacks that could compromise the integrity of the network. This section discusses how PoTS effectively mitigates threats such as Sybil attacks, 51\% attacks, and Distributed Denial of Service (DDoS) attacks.

In the case of Sybil attacks, where a malicious actor creates multiple fake identities to gain disproportionate influence over the network, PoTS employs a group-based approach that significantly reduces the impact of such attacks. Since groups in PoTS are formed randomly and require collaboration among members, it becomes exceedingly difficult for an attacker to control a significant portion of the network through fake identities alone.

Regarding 51\% attacks, where an attacker attempts to control the majority of the network's computational power, PoTS provides a layer of security by distributing computational tasks across multiple groups. This decentralization of power makes it challenging for any single entity to gain the majority control needed to carry out a 51\% attack. Even if an attacker were to infiltrate several groups, the collaborative nature of PoTS would require the attacker to compromise a substantial portion of the network, making such an attack highly impractical.

Lastly, against DDoS attacks, PoTS benefits from its decentralized structure. The distribution of computational tasks and decision-making processes across numerous groups reduces the likelihood that any single point of failure could disrupt the entire network. The network's resilience is further enhanced by the random and dynamic nature of group formation, which ensures that even if one group is targeted, the network as a whole can continue to function effectively.

Overall, PoTS's design incorporates multiple layers of security that protect the network from common threats, ensuring that it can offer a level of security comparable to or even exceeding that of traditional PoW systems.

\paragraph{Role of Randomness and Group Dynamics}
Randomness and group dynamics play a crucial role in enhancing the security of PoTS. The randomness in group formation ensures that participants are randomly assigned to groups, making it difficult for attackers to predict or influence the composition of any given group. This unpredictability acts as a deterrent against coordinated attacks, as attackers cannot easily position themselves within the network to maximize their influence.

Additionally, the dynamics within each group contribute to the overall security of the network. Since each group is composed of participants who may not have prior associations, the necessity of collaboration to achieve a common goal---solving the cryptographic puzzle---forces participants to work together. This collaborative environment further complicates the efforts of any attacker attempting to dominate or disrupt the network, as the group's success relies on the honest participation of its members.

The combination of randomness in group formation and the collaborative dynamics within groups creates a robust defense mechanism in PoTS. These elements ensure that the network remains resilient against attacks, as they introduce significant barriers to any malicious actor attempting to gain control or disrupt the network's operations. By leveraging these features, PoTS enhances the security of the blockchain, making it a formidable alternative to traditional consensus mechanisms.

\section{Energy Consumption Analysis}
\subsection{Assumptions and Definitions}
\subsubsection{Assumptions}
In this analysis, several simplifying assumptions are made to evaluate the theoretical upper bound of energy consumption:

\begin{itemize} 
\item In PoW, the computation by each participant is assumed to complete simultaneously, implying zero variance in computation time. However, it is assumed that one participant will finish first, determined by a negligible time difference. 
\item Network latency is assumed to be negligible, effectively set to zero. 
\item The computation in PoW is assumed to be executed sequentially by $N$ participants in PoTS, with the energy consumption distributed across the group. 
\item In PoTS, the rewards obtained are assumed to be equally divided among the $N$ participants in each group. 
\item When not performing work for either PoW or PoTS, the energy consumption resulting from these algorithms is assumed to be zero. It is considered that the participants may be engaged in other useful tasks. 
\end{itemize}

\subsubsection{Definition of Variables:}
The following variables are defined for the analysis:
\begin{itemize}
\item $n$: the total number of participants in the network.
\item $N$: the number of participants per group.
\item $M$: the number of groups, that is $n = M \times N$, and thus $M = n / N$.
\item $E_W$: the energy consumption per participant in PoW.
\item $R_W$: the reward obtained by the first participant to complete the computation in PoW.

\end{itemize}

\begin{figure}[t]
\begin{center}
\includegraphics[scale=0.5]{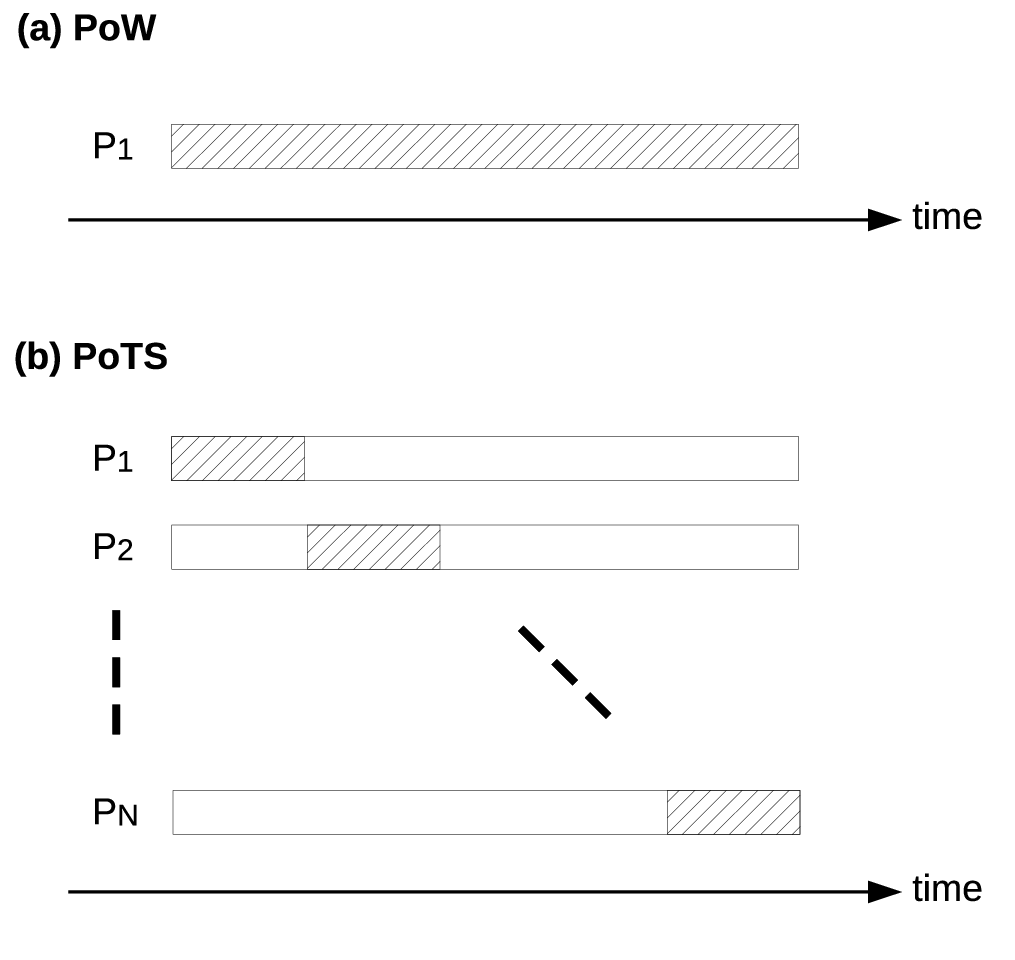}
\caption{Energy Consumption Comparison between PoW and PoTS}
\label{fig:pow-pots-time-diagram}
\end{center}
\end{figure}

\subsection{Results and Analysis}
Figure \ref{fig:pow-pots-time-diagram} illustrates the process of computation in both PoW and PoTS. The horizontal axis represents time.
In Figure \ref{fig:pow-pots-time-diagram} (a), it shows the time taken by a single participant in PoW to perform computations, where the busy state of the machine is indicated by shaded areas.
In Figure \ref{fig:pow-pots-time-diagram} (b), it depicts the busy state of a single group composed of $N$ participants in PoTS.
For the entire network, PoW involves computations by $n$ times the number of participants shown in the figure, while PoTS involves $M$ times the number of participants performing computations simultaneously.

\subsubsection{Energy Consumption in PoW}
In PoW, each participant independently consumes $E_W$ of energy, leading to a total network consumption of $n \times E_W$. The first participant to complete the computation receives the reward $R_W$.

\subsubsection{Energy Consumption in PoTS}
In PoTS, the energy consumption is distributed across the participants within each group. The energy consumption per group is $E_W$, consistent with the assumption that the computation in PoW is executed sequentially by $N$ participants in PoTS. Given that there are $M = n / N$ groups in the network, the total energy consumption for the entire network is $M \times E_W = n \times E_W / N$. This indicates that the energy consumption in PoTS is $1/N$ of that in PoW, demonstrating a significant improvement in energy efficiency.

\subsubsection{Comparative Analysis}
Comparing the energy consumption between PoW and PoTS reveals that PoTS can reduce the total energy consumption of the network by a factor of $1/N$. In PoTS, the reward is distributed equally among the $N$ participants within each group, resulting in an expected reward per participant that is equivalent to that in PoW, i.e., $R_W / n$\footnote{Specifically, the expected reward for each group in PoTS is $R_W/M = R_W \times N / n$, and since this reward is equally divided among the $N$ participants in the group, the expected reward per participant becomes $(R_W \times N / n) / N = R_W / n$.}. This analysis shows that PoTS not only improves energy efficiency but also maintains a fair incentive structure, ensuring that participants receive an equitable share of the rewards.

\section{Discussion}
\subsection{Implications for Blockchain Technology}
\subsubsection{Adoption of Energy-Efficient Consensus Mechanisms}
The adoption of energy-efficient consensus mechanisms is poised to have a profound impact on the future of blockchain technology. As demonstrated in Section 4, the PoTS algorithm significantly reduces the total energy consumption of the network compared to traditional PoW systems. Specifically, PoTS can reduce the energy consumption by a factor of $1/N$, where $N$ is the number of participants in each collaborative group. This reduction in energy use directly addresses one of the most pressing criticisms of blockchain technology: its substantial environmental impact.

The environmental concerns surrounding PoW-based blockchains have led to increasing scrutiny from both regulators and the public. The high energy demand of PoW, which rivals that of entire nations, poses a significant barrier to the broader adoption of blockchain technologies, especially in industries where sustainability is a critical consideration. By adopting energy-efficient mechanisms like PoTS, blockchain networks can drastically lower their carbon footprint, making them more sustainable and appealing to a wider range of stakeholders.

Moreover, the transition to energy-efficient consensus mechanisms does not merely offer environmental benefits. Maintaining network security while reducing energy consumption is paramount to ensuring the continued viability and scalability of blockchain systems. PoTS, with its collaborative approach, preserves the high level of security expected from blockchain networks by distributing the computational workload and enhancing the resilience of the system against attacks. This balance between energy efficiency and security is essential for the next phase of blockchain evolution, where both technological innovation and environmental stewardship must go hand in hand.

\subsubsection{PoTS as a Sustainable Alternative to PoW}
PoTS emerges as a compelling alternative to PoW, particularly from a sustainability perspective. As highlighted in the energy consumption analysis, PoTS not only reduces the total energy usage of blockchain networks but also maintains an equitable reward distribution among participants, ensuring that the incentives for network maintenance are preserved.

The sustainability of blockchain technology depends on its ability to minimize energy consumption without compromising security. PoTS achieves this by leveraging group collaboration, where participants work together to complete the necessary computations, rather than competing individually as in PoW. This collaborative approach results in a more efficient use of resources, reducing redundant calculations and lowering the overall energy demand. Consequently, PoTS can mitigate the environmental impact of blockchain networks, making them more viable in the long term.

Furthermore, the adoption of PoTS could drive broader acceptance of blockchain technology across various sectors. Industries that prioritize sustainability, such as renewable energy, supply chain management, and carbon trading, could particularly benefit from the energy efficiency offered by PoTS. By reducing the energy costs associated with maintaining blockchain networks, PoTS can also lower operational expenses, making blockchain a more attractive solution for a wide range of applications.

In summary, PoTS represents a sustainable alternative to PoW, offering a path forward for blockchain technology that aligns with global efforts to reduce energy consumption and combat climate change. As blockchain continues to evolve, the integration of energy-efficient consensus mechanisms like PoTS will be crucial in ensuring that this transformative technology can meet the needs of a rapidly changing world.

\subsection{Limitations and Challenges}
\subsubsection{Scalability Concerns}
While PoTS offers significant improvements in energy efficiency and sustainability, its scalability in large-scale networks presents certain challenges that need to be addressed. One of the primary concerns is how PoTS will function as the network size increases. As the number of participants ($n$) grows, the complexity of coordinating and managing the groups of $N$ participants may become a bottleneck. Ensuring that each group functions effectively without introducing delays or inefficiencies is crucial for maintaining overall network performance.

In large-scale networks, the distribution of computational load across multiple participants is intended to enhance efficiency. However, as the network expands, the coordination overhead could potentially negate the benefits of this distributed approach. For example, the process of forming and reconfiguring groups dynamically, especially in environments where nodes frequently join or leave the network, could introduce significant latency and reduce the system's responsiveness. Additionally, as $N$ increases, the likelihood of network fragmentation or uneven distribution of computational power across groups could lead to inefficiencies and security vulnerabilities.

To address these scalability concerns, several strategies could be explored. One potential solution is the implementation of more sophisticated group formation algorithms that optimize for both efficiency and security, even as the network scales. Another approach could involve hierarchical or layered network structures, where groups are organized in a multi-tiered system to manage complexity and reduce the coordination burden on any single group. Moreover, adaptive mechanisms that adjust the size of $N$ based on real-time network conditions could help maintain optimal performance across varying network sizes. By anticipating and addressing these challenges, PoTS can be made more scalable and robust, enabling it to function effectively in networks of all sizes.

\subsubsection{Potential Barriers to Adoption}
Despite the advantages of PoTS in terms of energy efficiency and sustainability, there are several potential barriers to its widespread adoption. One of the key challenges is the technical complexity involved in transitioning from existing PoW-based systems to PoTS. Many current blockchain networks are deeply entrenched in PoW, and migrating to a new consensus mechanism would require significant changes to the underlying infrastructure. This includes the need for extensive testing, validation, and the development of new tools and protocols to support the PoTS framework.

Another barrier is the design of incentives that encourage participants to cooperate within the PoTS framework. Unlike PoW, where individual competition drives participation, PoTS relies on collaborative efforts within groups. Designing an incentive structure that not only motivates participants to contribute fairly but also prevents freeloading or collusion is a complex task. Ensuring that all participants receive equitable rewards while maintaining the security and integrity of the network is crucial for the successful adoption of PoTS.

Furthermore, the adoption of any new consensus mechanism requires building trust among stakeholders. PoTS must demonstrate its reliability, security, and performance in real-world applications before it can gain the confidence of the broader blockchain community. Overcoming skepticism, particularly in industries where stability and security are paramount, will be essential. This may involve extensive outreach, education, and the establishment of pilot projects that showcase the benefits of PoTS in a controlled environment.

To overcome these barriers, a phased approach to adoption may be necessary. Initial deployments of PoTS in smaller, less critical networks could serve as testbeds to refine the mechanism and build a track record of success. Gradual scaling, coupled with continuous feedback and improvements, would help in gaining the trust of the community and ensuring a smoother transition from PoW to PoTS. Addressing these challenges proactively is key to realizing the full potential of PoTS as a sustainable and scalable consensus mechanism for the future of blockchain technology.

\section{Conclusion}

In conclusion, this research has highlighted the significant advantages of Proof of Team Sprint (PoTS) as an alternative to Proof of Work (PoW). PoTS is a consensus algorithm that organizes participants into $M$ groups of $N$ members, who collaboratively work to solve cryptographic puzzles. Each group functions as a team, distributing the computational workload among its members, which allows the network to achieve the same security as PoW but with significantly reduced energy consumption. 
The findings, representing the theoretical upper bound of energy consumption, demonstrate that PoTS not only reduces energy consumption by a factor of $1/N$, but also maintains a fair and equitable reward distribution among participants.
This addresses two of the major challenges faced by current blockchain systems: high energy consumption and reward fairness. These results underscore the potential of PoTS to contribute meaningfully to the advancement of blockchain technology by offering a more sustainable and efficient consensus mechanism.

The contributions of this research to the broader field of blockchain are substantial. By providing a new approach to consensus that balances energy efficiency with security and fairness, PoTS represents a critical step forward in the development of sustainable blockchain technologies. This work lays the foundation for future advancements that could further reduce the environmental impact of blockchain while enhancing its applicability to a wider range of use cases.

As we look towards the future, several areas of development are critical for enhancing the security and scalability of the PoTS algorithm. While PoTS has demonstrated significant potential in terms of energy efficiency and fair reward distribution, there are challenges that must be addressed to ensure its viability in larger and more complex networks. Future research should focus on strengthening the security mechanisms of PoTS, particularly in environments with a high number of participants. This could involve exploring advanced cryptographic techniques to protect against potential attacks and ensuring that the collaborative nature of PoTS does not introduce vulnerabilities that could be exploited.

Scalability is another key area where PoTS will need to evolve. As networks grow, the ability to efficiently manage and coordinate large groups of participants becomes increasingly important. Research into more sophisticated group formation algorithms, as well as hierarchical structures that can distribute the computational load more effectively, will be essential for maintaining PoTS's performance in large-scale networks. Addressing these scalability concerns will be crucial for PoTS to achieve widespread adoption across various blockchain platforms.

Beyond its application in blockchain technology, PoTS's energy-efficient and cooperative consensus mechanism holds promise for other areas of distributed computing. The principles underlying PoTS could be adapted to improve the efficiency of other distributed systems, particularly in contexts where energy consumption is a significant concern. For example, PoTS could be applied to environmental monitoring networks, smart grids, or any system where reducing energy use while maintaining robust security is paramount. Expanding PoTS to these new domains could not only extend its utility but also amplify its impact on sustainability and resource management.

In addition to these applications, further optimization of PoTS's energy consumption remains a priority. While the current implementation offers substantial energy savings compared to traditional PoW, there is always room for improvement. Future work should explore refinements to the PoTS algorithm that could reduce energy consumption even further, possibly through more efficient hardware implementations or innovative energy management techniques. By continuing to optimize PoTS, we can ensure that it remains at the forefront of sustainable blockchain technology.

Looking ahead, the future of PoTS appears promising. As the technology matures, PoTS has the potential to contribute meaningfully to the evolution of blockchain systems. Continued research and development will be important in realizing its potential, particularly in addressing challenges related to security and scalability in larger networks. Additionally, while PoTS may find applications beyond blockchain, its impact on global sustainability efforts remains an area for further exploration. The progress towards broader adoption and optimization of PoTS is still in its early stages, and its future applications hold considerable promise.

\bibliographystyle{IEEEtran}
\bibliography{pots}

\end{document}